\documentclass[fleqn,usenatbib]{mnras}
\usepackage{savesym}
\usepackage{amsmath}
\savesymbol{iint}
\usepackage{txfonts}
\restoresymbol{TXF}{iint}

\usepackage[T1]{fontenc}
\usepackage{ae,aecompl}

\usepackage{graphicx}	
\usepackage{amsmath}	
\usepackage{amssymb}	

\title[System mass constraints for the AMXP XTE J1814-338]{System mass constraints for the accreting millisecond pulsar XTE J1814-338 using Bowen fluorescence}

\author[L. Wang et al.]{
L. Wang,$^{1}$\thanks{E-mail: \href{mailto:Zhuqing.Wang@warwick.ac.uk}{Zhuqing.Wang@warwick.ac.uk}}
D. Steeghs,$^{1}$
J. Casares,$^{2, 3, 4}$
P. A. Charles,$^{5, 6}$
T. Mu\~noz-Darias,$^{2, 3}$
\newauthor
\:T. R. Marsh,$^{1}$
R. I. Hynes$^{7}$ and
K. O'Brien$^{4}$
\\
$^{1}$Department of Physics, University of Warwick, Gibbet Hill Road, Coventry CV4 7AL, UK\\
$^{2}$ Instituto de Astrof\'isica de Canarias, 38205 La Laguna, Tenerife, Spain\\
$^{3}$ Departamento de astrof\'isica, Univ. de La Laguna, E-38206 La Laguna, Tenerife, Spain\\
$^{4}$ Department of Physics, Astrophysics, University of Oxford, Denys Wilkinson Building, Keble Road, Oxford OX1 3RH, UK\\
$^{5}$ Dept of Physics \& Astronomy, University of Southampton, Southampton SO17 1BJ, UK\\ 
$^{6}$ Astrophysics, Cosmology and Gravity Centre (ACGC), University of Cape Town, Private Bag X3, Rondebosch, 7701, South Africa\\ 
$^{7}$ Department of Physics and Astronomy, Louisiana State University, Baton Rouge, LA 70803, USA\\
}

\date{Accepted XXX. Received YYY; in original form ZZZ}

\pubyear{2015}

\begin{document}
\label{firstpage}
\pagerange{\pageref{firstpage}--\pageref{lastpage}}
\maketitle

\begin{abstract}
We present phase-resolved spectroscopy of the millisecond X-ray pulsar XTE J1814-338 obtained during its 2003 outburst.
The spectra are dominated by high-excitation emission lines of \ion{He}{ii} $\lambda$4686, H$\beta$, and the Bowen blend \ion{C}{iii}/\ion{N}{iii} 4630-50\AA. We exploit the proven Bowen fluorescence technique to establish a complete set of dynamical system parameter constraints using \textit{bootstrap Doppler tomography}, a first for an accreting millisecond X-ray pulsar binary. The reconstructed Doppler map of the \ion{N}{iii} $\lambda$4640 Bowen transition exhibits a statistically significant (> 4$\sigma$) spot feature at the expected position of the companion star.  
If this feature is driven by irradiation of the surface of the Roche lobe filling companion, we derive 
a strict lower limit to the true radial velocity semi-amplitude $K_2$.
Combining our donor constraint with the well constrained orbit of the neutron star leads to a determination of the binary mass ratio: $q$ = $0.123^{+0.012}_{-0.010}$. 
The component masses are not tightly constrained given our lack of knowledge of the binary inclination. 
We cannot rule out a canonical neutron star mass of 1.4 $M_{\odot}$ (1.1 $M_{\odot}$ $<$ $M_1$ $<$ 3.1 $M_{\odot}$; 95\%). 
The 68/95\% confidence limits of $M_2$ are consistent with the companion being a significantly bloated, M-type main sequence star.
Our findings, combined with results from studies of the quiescent optical counterpart of XTE J1814-338, suggest the presence of a rotation-powered millisecond pulsar in XTE J1814-338 during an X-ray quiescent state. The companion mass is typical of the so-called `redback' pulsar binary systems ($M_2$ $\sim$ 0.2 $M_{\odot}$). 

\end{abstract}

\begin{keywords}
accretion, accretion discs -- binaries: close -- stars: individual: XTE J1814-338 -- X-rays: stars
\end{keywords}



\section{Introduction}


Low-mass X-ray binaries (LMXBs) are systems in which a neutron star (NS) or a black hole (BH) accretes matter from a low-mass companion star. 
Since the discovery of the first millisecond radio pulsar \citep[MSP;][]{1982Natur.300..615B}, it has been suspected that long periods of mass transfer onto old NSs hosted in LMXBs might be responsible for spinning up the compact object to the ms regime. 
 A `recycled' MSP is thought to be formed when accretion turns off completely \citep{1982Natur.300..728A}. 
The detection of the first accreting millisecond X-ray pulsar (AMXP) SAX J1808.4-3658 in the course of an X-ray outburst episode \citep{1998Natur.394..344W} provided a nice confirmation of the recycling scenario.
The radio pulsar/LMXB link was firmly confirmed with more recent discoveries of transitional millisecond pulsar binaries. The most notable examples include the `missing link pulsar' PSR J1023+0038 that turned on as an MSP after a LMXB phase \citep{2009Sci...324.1411A}; and the direct evolutionary link IGR J18245-2452, which has shown both an MSP and an AMXP phase \citep{2013Natur.501..517P}; and XSS J1227.0-4859 (see, e.g., \citealt{2014MNRAS.444.3004D}; \citealt{2015ApJ...800L..12R}).

The fifth accreting millisecond X-ray pulsar XTE J1814-338 was discovered on June 5 2003 by the \textit{Rossi X-ray Timing Explorer (RXTE)} satellite during routine observations of the Galactic-center region, and has a pulse frequency of 314 Hz \citep{2003IAUC.8144....1M}. 
Among the $\sim$15 known AMXP systems (see \citealt{2012arXiv1206.2727P} for a recent review), its 4.3 hrs orbital period is the most similar to the period of `classic' non-pulsing neutron star LMXBs, e.g., 4U 1735-444 (4.7 hrs), GX 9+9 (4.2 hrs) or 4U 1636-536 (3.8 hrs).
A total of 28 type I X-ray bursts have been observed from XTE J1814-338 to date, all with burst oscillations in the vicinity of the 314 Hz pulsar frequency, confirming that the burst oscillation frequency for XTE J1814-338 is at the NS spin frequency (see, e.g., \citealt{2005ApJ...634..547W}). A source distance of 8 $\pm$ 1.6 kpc was inferred from the last burst which likely reached the Eddington luminosity \citep{2003ApJ...596L..67S}.   

The presence of ms pulsations provides the opportunity for a precise determination of the orbit of the neutron star \citep{2007MNRAS.375..971P}. However, in order to establish a complete set of system parameters, including the mass of the neutron star, 
the radial velocity curve of the companion is required. 
Optical and near-infrared observations of AMXPs in quiescence can offer promising opportunities to constrain the radial velocity of the donor, as the faint donor star may be detected directly. But so far, these have only led to upper limits or a counterpart that is too faint on which to perform radial velocity studies \citep[e.g.][]{2001MNRAS.325.1471H,2003MNRAS.344..201J, 2005ApJ...627..910K}. 

A different avenue for determining system parameters of LMXBs and/or AMXPs \textit{in active states} was opened up by \citet{2002ApJ...568..273S}. 
High resolution, phase-resolved blue spectroscopy of the prototypical LMXB Scorpius X-1 revealed extremely narrow, high-excitation emission components arising from the surface of the donor star, leading to the first radial velocity curve for the mass donor in Sco X-1 and binary parameter constraints in support of the presence of a 1.4 $M_{\odot}$ neutron star. 
These narrow emission lines were strongest in the Bowen region (4630-4650 $\AA$) consisting of a blend of \ion{N}{iii} and \ion{C}{iii} lines, and are the result of fluorescence of the gas by UV photons from the hot inner disc \citep{1975ApJ...198..641M}. 
This technique has been used to constrain the orbital parameters of a number of persistent LMXBs (see \citealt{2008AIPC.1010..148C} for a review) as well as transient systems during their outbursts, e.g. GX 339-4 \citep{2003ApJ...583L..95H}, SAX J1808.4-3658 \citep{2009A&A...495L...1C}.

In this study we present medium resolution blue spectroscopy of the accretion-driven millisecond pulsar XTE J1814-338 obtained with the VLT during its 2003 outburst. 
The Bowen fluorescence technique is revisited and applied to XTE J1814-338 to establish dynamical system parameter constraints, which can offer insights into the evolutionary scenario involving binary pulsars. Section \ref{sec:obs} summarizes the observing strategy and data reduction steps. 
In section \ref{sec:ave_spec} we present the average spectrum and main emission line parameters while radial velocities of emission lines are presented in section \ref{sec:rv}.
In section \ref{sec:method}, we exploit the Doppler tomography technique and further develop our methodology to obtain robust binary parameter constraints. Estimation of component masses and discussion of the results are given in section \ref{sec:discussion}.

\section{Observations and data reduction}
\label{sec:obs}
We observed XTE J1814-338 (hereafter, J1814) using the FORS2 Spectrograph attached to the 8.2m Yepun Telescope (UT4) at Observatorio Monte Paranal (ESO) on the night of 23 June 2003 (programme 071.D-0372). 
A total of twenty spectra were obtained with the R1400V holographic grating, covering a complete orbital cycle using 700s long exposures. Matching with an 0.7 arcsec slit width resulted in a wavelength coverage of 4514-5815 $\AA$ at 70 km s$^{-1}$ (FWHM) resolution.  The seeing was variable between 0.6$\arcsec$ -- 1.2$\arcsec$ during our run. 
The slit position angle was fixed to PA=100$\degr$ which coincides with the parallactic angle at the end of our run, when the airmass of the target was highest (sec$z$ = 2.1). At the same time, two comparison stars were included in the slit, which enabled us to monitor slit losses and obtain a relative flux calibration. 
The flux standard Feige 110 was also observed with the same instrumental configuration to correct for the instrumental response of the detector.

The images were de-biased and flat-fielded, and the spectra were subsequently extracted using conventional optimal extraction techniques in order to optimize the signal-to-noise ratio of the output \citep{1986PASP...98..609H}.
A He+Ne+Hg+Cd comparison lamp image was obtained in the daytime to provide the wavelength calibration scale. A 4th-order polynomial fit to 19 arc lines produced an rms scatter $<$ 0.05 $\AA$ with a mean dispersion of 0.64 \AA~pix$^{-1}$. 
Instrumental flexure was monitored through cross-correlation between the sky spectra and was found to be very small (always within 5 km s$^{-1}$). These velocity drifts were nevertheless removed from each individual spectra, and the zero point of the final wavelength scale was established by the strong OI sky line at 5577.338 $\AA$.  
Individual spectra were corrected for slit losses using the brighter comparison star that shared the slit with our target. 
Correction factors were computed using a low-order spline fit to the ratio between each comparison star spectrum and its grand sum average. The target spectra were then divided by these fits to correct for time-dependent slit losses. 


\newcommand{\kms}{\,km\,s$^{-1}$}
\section{Average spectrum and orbital variability}
\label{sec:ave_spec}

\begin{figure*}
	\includegraphics[width=2.0\columnwidth]{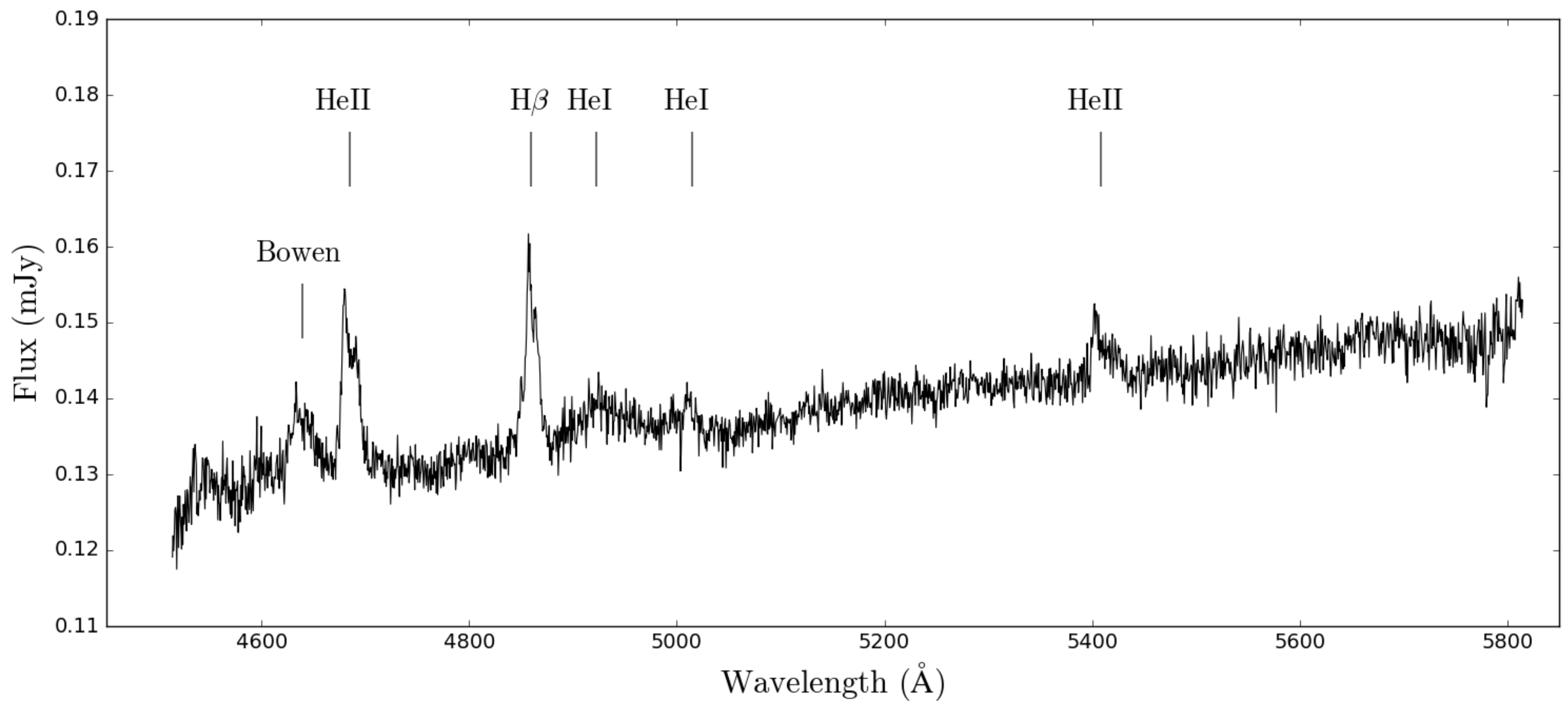}
    \caption{The average optical spectrum of XTE J1814-338. Main emission lines are indicated.}
    \label{fig:spec}
\end{figure*}

The average spectrum of J1814 is presented in Fig.~\ref{fig:spec}.
The spectrum is dominated by high excitation emission lines of \ion{He}{ii} $\lambda\lambda$4686 \& 5411$\AA$, the Bowen blend at $\lambda$$\lambda$4630-50 and H$\beta$, which are typical of LMXBs and Soft X-ray Transients (SXTs) in outburst and first noted in \cite{2003IAUC.8155....2S}. 
\ion{He}{i} lines at $\lambda$4922 and $\lambda$5015 are significantly weaker and no absorption components are discernible in the Balmer or \ion{He}{i} lines, as seen in high inclination systems such as X1822-371 \citep[see][]{2003ApJ...590.1041C}. 
The \ion{He}{ii} profiles are double-peaked, as expected from an accretion disc, but the blue peak is considerably stronger than the red peak.  The H$\beta$ profile is more complex and its blue side peaks at lower velocities. A list of time-averaged emission line parameters is presented in Table~\ref{tab:lineparams}. 

\begin{table}
	\centering
	\caption{Emission line parameters.}
	\label{tab:lineparams}
	\begin{tabular}{lccc} 
		\hline
		Line & Centroid & FWHM & EW\\
                & {\AA} & {km s$^{-1}$} & {\AA} \\
		\hline
                Bowen    & 4639.0 $\pm$ 0.8 & 1613 $\pm$ 180 & 2.2 $\pm$ 0.2 \\ 
                \ion{He}{ii} & 4684.6 $\pm$ 0.4 & 1160 $\pm$  70 & 2.9 $\pm$ 0.2 \\
                H$\beta$ & 4859.7 $\pm$ 0.3 &  920 $\pm$  60 & 2.8 $\pm$ 0.2  \\
                \ion{He}{ii} & 5408.5 $\pm$ 1.7 & 1190 $\pm$ 200 & 0.8 $\pm$ 0.2  \\ 
		\hline
	\end{tabular}
\end{table}

\begin{figure}
	\includegraphics[width=\columnwidth]{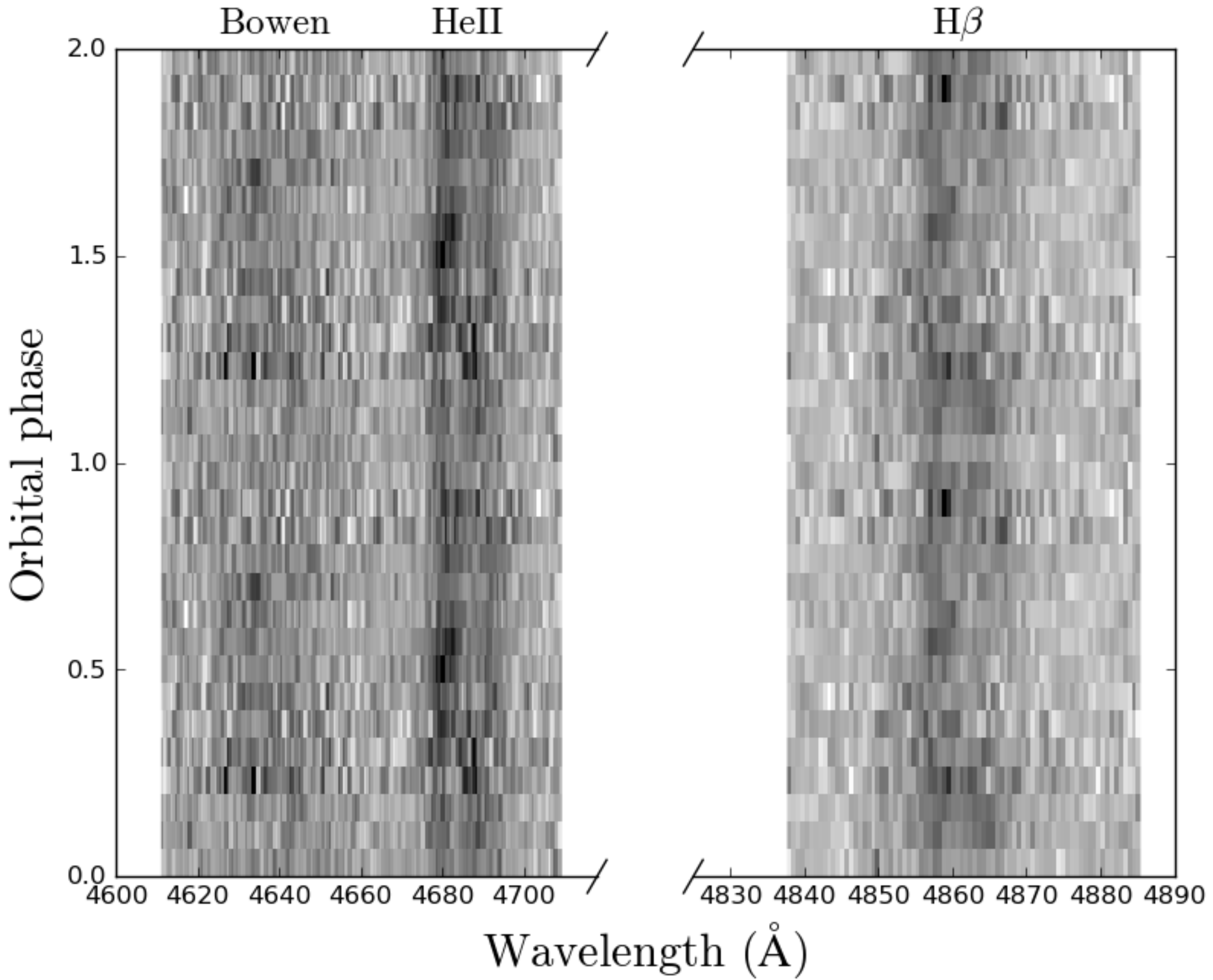}
    \caption{Trailed spectra showing the orbital evolution of the strongest emission lines in 15 phase bins.}
    \label{fig:trail}
\end{figure}

Fig.~\ref{fig:trail} displays the orbital evolution of the most intense emission features presented in 15 phase bins. Binary phases were computed using the extremely accurate X-ray pulsar orbital solution \citep{2007MNRAS.375..971P} :
\begin{equation}
    T_{0} $\rm~HJD(UTC)$ =2452798.3539536(9)+0.178110219(2) E.
	\label{eq:ephemeris}
\end{equation}

\noindent
Here we adjusted $T_{0}$ such that phase 0 corresponds to the inferior conjunction of the companion star in a heliocentric UTC time system. \ion{He}{ii} $\lambda$4686 shows a clear double-peaked profile with the blue peak stronger around phase $\sim$ 0.6. An S-wave is visible, crossing from blue to red velocities at phase $\sim$ 0.9. The H$\beta$ profiles are dominated by the blue peak throughout the orbit whereas the Bowen blend is too noisy to directly reveal any multi-component structure in the raw data. In general, the line intensities appear to peak in flux around phase $\sim$ 0.3 - 0.8.

\section{Radial velocities and the systemic velocity}
\label{sec:rv}

\begin{figure}
	\includegraphics[width=\columnwidth]{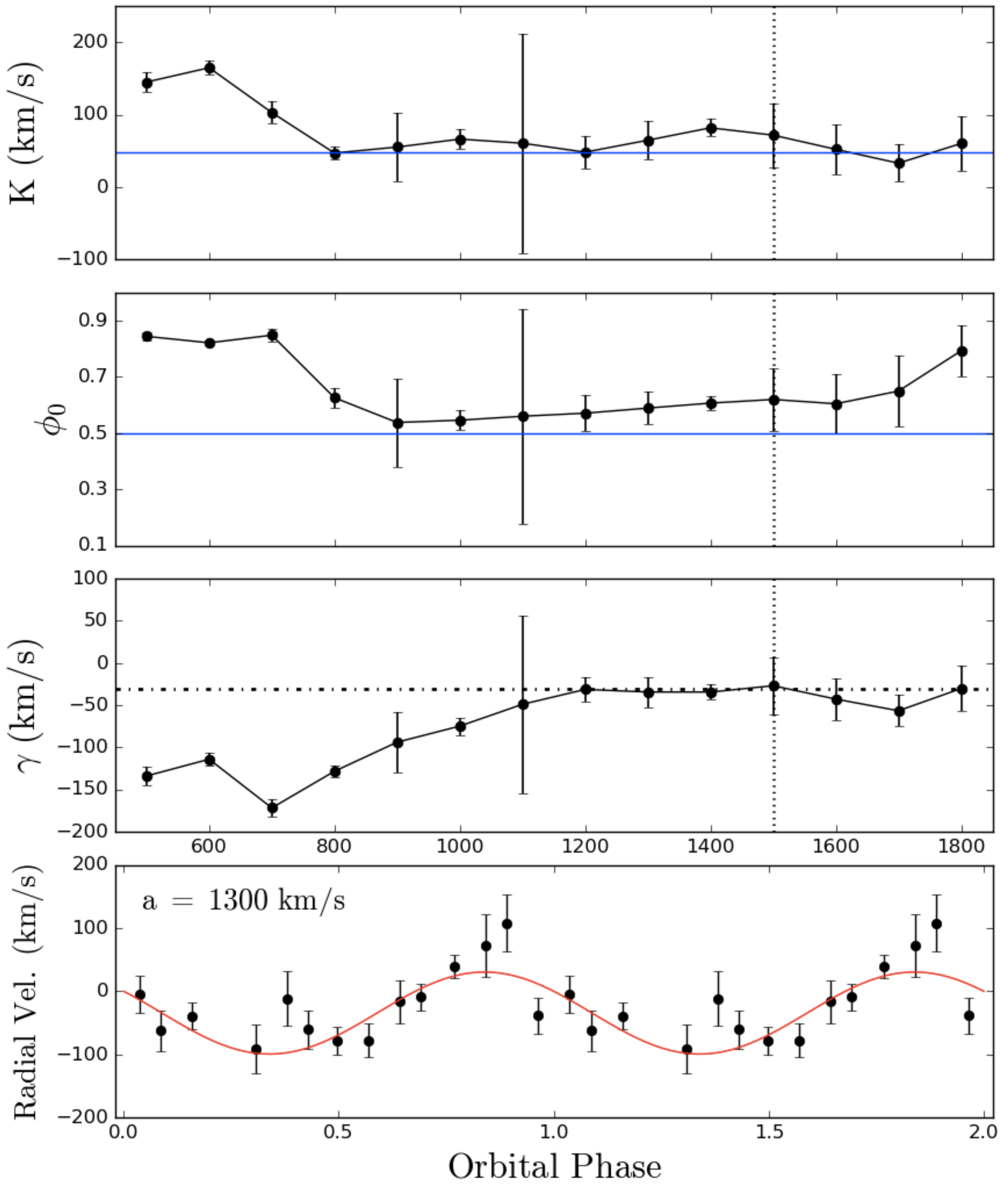}
    \caption{The diagnostic diagram of \ion{He}{ii} $\lambda$4686 (top 3 panels). Fitted sine-wave parameters (with 1$\sigma$ errors) are plotted as functions of the Gaussian separation $a$. The vertical dotted line denotes a limit of the Gaussian separation above which the continuum noise dominates the radial velocities. The bottom panel shows the radial velocity curve for $a$ = 1300 km s$^{-1}$ overplotted with the best-fit sine wave (red curve).}
    \label{fig:shafter_diagram}
\end{figure}

The broad disc emission features may be used to derive radial velocity curves as the disc gas should trace Keplerian orbits around the neutron star. In order to avoid contamination from line core components (see Fig.~\ref{fig:trail}) we have applied the well established double-Gaussian technique \citep{1980ApJ...238..946S} to \ion{He}{ii} $\lambda$4686. The Gaussian widths were fixed to FWHM = 200 km s$^{-1}$ with a variable separation $a$ = 500 - 1800 km s$^{-1}$ in steps of 100 km s$^{-1}$.  The radial velocity (RV) curves were fitted with a sine function and the resultant parameters are displayed using the traditional diagnostic diagram 
\citep[Fig.~\ref{fig:shafter_diagram};][]{1985ASSL..113..355S}.
As we move away from the line core and thus mainly probe the dynamics of the inner disc gas, the $\gamma$-velocity stabilizes towards -30 km s$^{-1}$ (horizontal dash-dotted line). The radial velocity amplitude K drops from $\sim$ 150 to 50 km s$^{-1}$, and the phasing decreases from $\sim$ 0.85 to 0.5-0.6.
From the pulsar solution we know the absolute phase (0.5) and velocity semi-amplitude of the neutron star ($K_1=47.848 \pm 0.001$ km s$^{-1}$, marked as horizontal blue lines in Fig.~\ref{fig:shafter_diagram}). 
It thus appears that in the range $a$ $\sim$ 900 - 1200 km s$^{-1}$, the disc dynamics tracks the true orbit of the neutron star quite well. At larger separations, parameters deviate again due to increasing noise as we probe the faint line wings.  Although we know $K_1$ already, this estimate suggests a systemic velocity near -30 km s$^{-1}$. 
In the next section, we present an alternative derivation of the systemic velocity  (see section \ref{sec:gamma}) using a new methodology, to be compared with the more traditional double Gaussian probe.

\section{Analysis}
\label{sec:method}
\subsection{Doppler mapping}
\label{sec:dmaps}
The discovery of narrow, high-excitation Bowen emission lines arising from the irradiated donor star in Sco X-1 opened up a new window to achieve robust radial velocity studies in luminous LMXBs \citep{2002ApJ...568..273S}.
Only in very few cases was it possible to use the conventional RV fitting method. 
For nearly all the fainter sources, the use of Doppler-tomography based methods was required in order to detect Bowen emission from the donor at low SNRs (S/N < 10). 

The Doppler tomography technique essentially uses all input spectra at once (ideally covering a whole orbital period) to invert phase-resolved data into an equivalent image of brightness distribution in velocity space or Doppler coordinates \citep{1988MNRAS.235..269M}.
This approach is well suited for faint donor components and has the advantage of being able to separate various sources of emission nicely in the output velocity-space image.
In the Doppler coordinate frame, the $V_x$-axis is defined by the direction from the accreting compact object to the donor star, the $V_y$-axis points in the direction of motion of the donor. If the input $\gamma$ parameter is close to the true systemic velocity, the origin at ($V_x$, $V_y$) = (0, 0) should then correspond to the centre of mass of the system. 
The transformation preserves the shape of the Roche lobe, and donor emission will always be mapped to a compact spot on the positive$\rm~V_{y}$-axis (provided that the correct ephemeris is used). 
If a Bowen spot is found at the expected position of the donor in the reconstructed tomogram, the (apparent) radial velocity semi-amplitude$\rm~(K_{em}$) of the secondary can be measured accurately from the map through a 2-dimensional Gaussian fit to the spot region. 

\begin{figure}
	\includegraphics[width=\columnwidth]{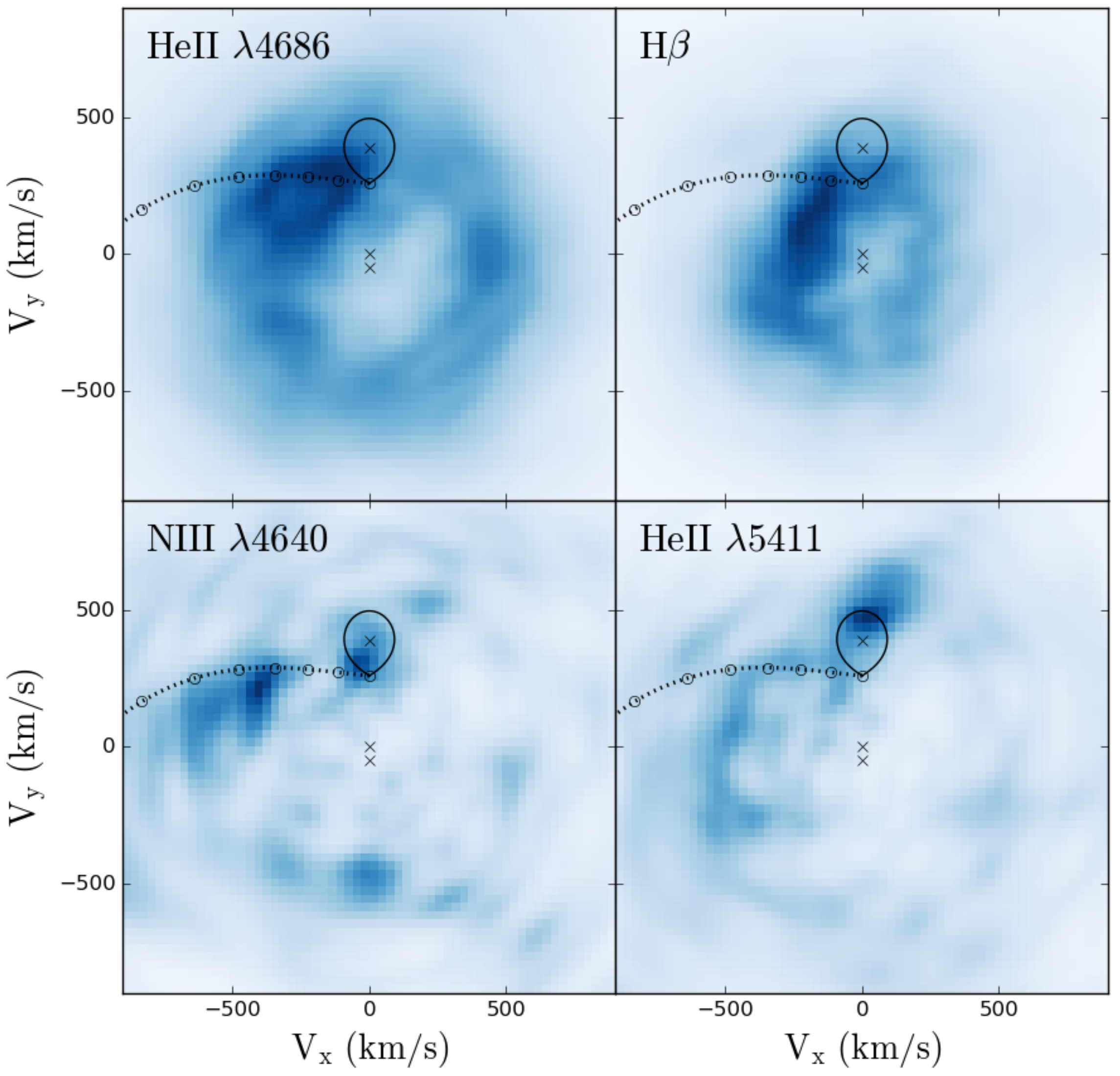}
    \caption{Doppler maps of several spectral features. All images were computed for a systemic velocity $\gamma$ = -30 km s$^{-1}$. We plot the gas stream trajectory and the Roche lobe of the donor star using $K_1$ = 47.848 km s$^{-1}$ and our best estimate of the mass ratio of 0.123 (see section \ref{sec:parameters}). The centre-of-mass velocity of the mass donor (0, $K_2$ = 390 km s$^{-1}$), the center of mass of the system (0, 0) and the velocity of the neutron star (0, -$K_1$) are denoted by crosses.}
    \label{fig:dopmaps}
\end{figure}

For J1814, we first prepared the spectra by subtracting a low order spline fit to the continuum regions, and subsequently used the second generation (Python$/$C++ based), maximum entropy Doppler tomography code\footnote{\url{https://github.com/trmrsh/trm-doppler}} developed by T. Marsh \citep{2016MNRAS.455.4467M} to exploit all 20 spectra simultaneously. 
In Fig.~\ref{fig:dopmaps}, we show the Doppler tomograms of the principal lines seen in our spectra using the ephemeris in equation~(\ref{eq:ephemeris}) and $\gamma$ = -30 km s$^{-1}$. 

Both \ion{He}{ii} and H$\beta$ maps show the classic ring-like features corresponding to emission from the accretion disc. An extended bright spot can be seen in the upper left velocity quadrant of these maps, suggesting excited emission from the stream-disc impact region.
In addition to evidence of bright spot emission, the map of \ion{N}{iii} $\lambda$4640 (usually the strongest Bowen component) for J1814 reveals a sharp spot along the positive$\rm~V_{y}$-axis. The centroid is at the position $V_x$ $\sim$ -20 km s$^{-1}$ and $V_y$ $\sim$ 319 km s$^{-1}$. 
The shape and phasing of this spot feature suggest that we may have detected a signal from the irradiated companion.
Finally, we notice another bright spot at approximately the expected phase of the companion in the \ion{He}{ii} $\lambda$5411 map, the spot is clearly at a much larger velocity ($V_x$ $\sim$ 30 km s$^{-1}$, $V_y$ $\sim$ 483 km s$^{-1}$) and more extended than the \ion{N}{iii} $\lambda$4640 spot.

\subsection{Bootstrap Monte-Carlo}

\subsubsection{Significance testing}
Perhaps the most important step towards robust binary parameter estimation using the Bowen technique involves 
confirming the statistical significance of any donor star components and to determine their position in velocity space robustly.
This becomes increasingly difficult as we push into lower SNR regimes, with correspondingly noisier Doppler maps that may contain spurious features. 
Therefore, it is vital to test the significance of the features seen in the \ion{N}{iii} $\lambda$4640 and \ion{He}{ii} $\lambda$5411 maps of J1814, before carrying out further rigorous analysis based on our reconstructions.

After testing on both synthetic and real datasets, we have opted to use the robust and versatile bootstrap technique to derive estimates of the standard deviation and confidence intervals on a real dataset with unknown probability distribution function \citep{2001MNRAS.326...67W, 2015MNRAS.447..149L}. 
The idea of the bootstrap method is to use the data themselves as an estimator of their underlying parent distribution \citep{tEFR93a}.
By randomly selecting from the original dataset 
allowing duplicates, simulated datasets of the same size as the original sample can be created. 
To allow for approximately the same noise level to be present in the bootstrap dataset as in the original data, only the error bars of the observed data were manipulated \citep[see details in][]{2001MNRAS.326...67W}.
For each bootstrap dataset, we compute a reconstructed image (automated processes; see \ref{sec:boot}) in the same manner as for the original image.   

\subsubsection{Monte-Carlo Doppler mapping}
\label{sec:boot} 
Upon creating a large number of new maps by resampling from the original trailed spectra, histograms of spot parameters can be constructed from ensembles of 2D Gaussian fits. The histograms provide information about the unknown sampling distribution of each property, including the centroid position, peak intensity, and FWHM. We can estimate the mean, the variance and confidence regions for any of these properties based on the bootstrap distribution. 

We have carefully examined sources of undesirable (random) variations that have high potential to distort the shape of the resulting bootstrap distribution and lead to inaccurate conclusions. 
While it may seem natural to fit all bootstrap spectra to the same user-defined goodness-of-fit level ($\chi^{2}$) as the observed spectra, it turns out that for maximum entropy (MEM) Doppler mapping, setting the same reduced $\chi^{2}$ to aim for 
can lead to 
broad entropy distribution across bootstraps, especially in low SNR regimes.
This means that in some cases, the bootstrap map looks simpler or smoother than the original map; 
in other cases we may end up with much noisier reconstructions, where the code has a hard time to reach the target $\chi^{2}$ and attempts to fit noise as well as signal. 
This issue can be resolved by iterating all maps towards constant entropy (S) level, as suggested in \citet{2001LNP...573....1M}. 

For each inversion, we first construct a smooth map by specifying a high initial value of $\chi^{2}$, and then decrease $C_{\mathrm{aim}}$ in small steps until the corresponding entropy reaches the desired value. The grid search is implemented through custom wrapper functions around the new DOPPLER routines. One effectively searches for the optimal MEM solution, so that the realism (measured by $\chi^{2}$) and simplicity (measured by S) can be ideally balanced for all simulated images. 
Variation can be further reduced by increasing the size of the computational bootstrap ($N_B$). We have tested that $N_B$ $\gtrsim$ 2000 yields robust results within a reasonable time budget (several hours/days of computation). In the rest of the analysis, all errors and significance levels are calculated from more than 2000 bootstrap samples with maps iterated to reach the same image entropy.

\subsubsection{Confirmation of the donor signature}
We start by estimating the significance of the most promising feature present in the \ion{N}{iii} $\lambda$4640 Doppler map as well as the statistical error of the phase shift relative to the pulsar ephemeris (see \ref{sec:dmaps}). It is known that the choice of the input parameter $\gamma$ can have an impact on the final Doppler image. In cases of high spectral resolution and high SNRs, an independent estimate of $\gamma$ can be obtained by reconstructing a series of maps with a range of $\gamma$'s, and searching for the one that yields the best fit to the data (minimal $\chi^{2}$), as well as an ideal, artifact-free image \citep{2003MNRAS.344..448S}.

In the case of J1814, we also generated a series of \ion{N}{iii} $\lambda$4640 maps with $\gamma$ varying from -200 to 200 km s$^{-1}$ in steps of 5 km s$^{-1}$, but found no pronounced minimum in $\chi^{2}$. 
Reduced $\chi^{2}$ remained low ($\chi^{2}_{\nu}$ $<$ 1.4) for $\gamma$ between -200 and 30 km s$^{-1}$. Based on this test, we conclude that the systemic velocity is indeed more likely to be negative, but that our SNR is too low to permit us to tightly constrain $\gamma$. 
Among our \ion{N}{iii} $\lambda$4640 maps, only a subset ($\gamma$ between -60 and 10 km s$^{-1}$) reveals a sharp, Gaussian-type spot feature along the positive $V_y$-axis, which we interpreted as a possible donor signature. Thus we choose to consider all reconstructions within this $\gamma$ range in the subsequent analysis.
To suppress the noise and enhance signal strength, we also created the \emph{combined Bowen Doppler maps}, for which we included both the \ion{N}{iii} $\lambda$4640 and the \ion{N}{iii} $\lambda$4634 components\footnote{Relevant Bowen components (other than the \ion{N}{iii} $\lambda$4640 line) may be identified by creating a trial Doppler corrected average spectrum in the rest frame of the donor (see Fig.~\ref{fig:spec_dop_corrected}).}. 
The combined Bowen map is almost identical in structure with the \ion{N}{iii} $\lambda$4640 map (see Fig.~\ref{fig:dopmaps}).
\begin{figure*}
	\includegraphics[width=2\columnwidth]{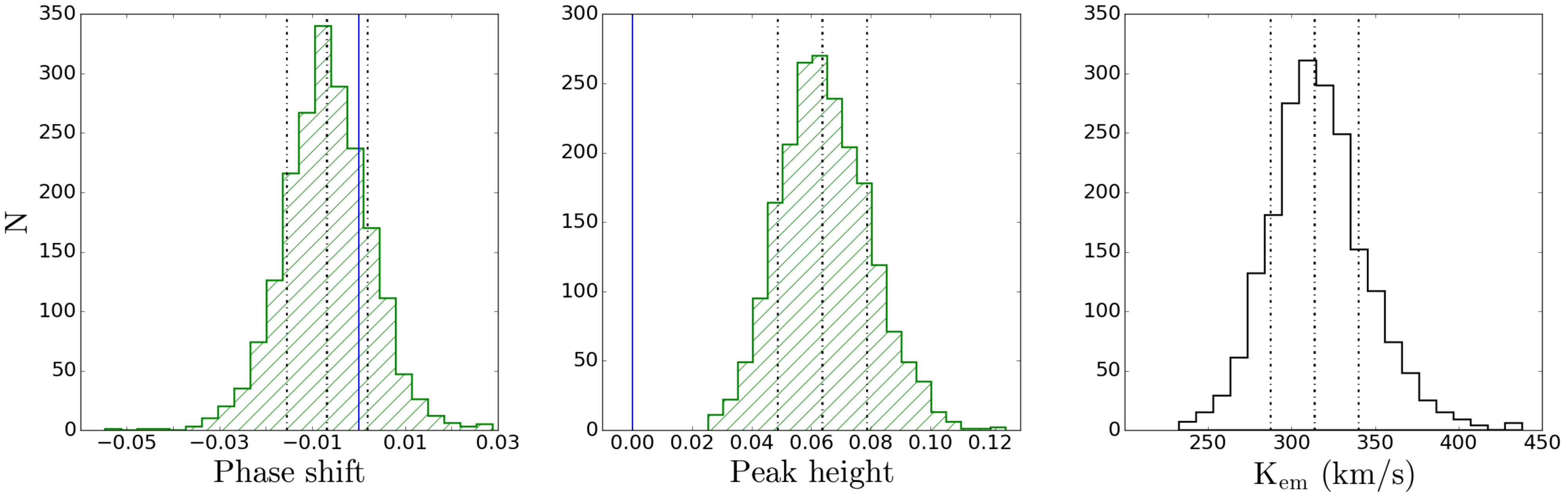}
    \caption{Number distributions of the phase shift (left), peak emission (middle) and radial velocity semi-amplitude (right) measured from 2000 bootstrap maps assuming a systemic velocity of -30 km s$^{-1}$. All maps were set to reach approximately the same entropy (S = -0.02). Dashed lines indicate the mean and the $\pm$ 1$\sigma$ confidence intervals. The phase shift relative to the pulsar ephemeris is consistent with zero (blue solid line), and the emission feature is significant at a $\sim$4.2$\sigma$ level.}
    \label{fig:histogram}
\end{figure*}
In Fig.~\ref{fig:histogram}, we give examples of histograms, for a $\gamma$ = -30 km s$^{-1}$ case, representing the bootstrap distribution of three important spot properties obtained from the \textit{combined} Bowen map.
 
Firstly, the bootstrap test allowed us to derive a phase shift (with its 1$\sigma$ statistical error) of $\Delta$$\mathrm{\phi_{spot}}$ = -0.007 $\pm$ 0.009 (-2.5 $\pm$ 3.1$^{\circ}$) using the centroid position ($V_x$, $V_y$). 
Given that the binary ephemeris is accurately known thanks to the pulsar, we can thus formally demonstrate that the sharp Bowen spot is effectively centred on the $V_y$-axis, exactly where the donor star should appear. 

Secondly, since all the distributions we obtained (see, e.g., Fig.~\ref{fig:histogram}) were Gaussian, we estimated the mean and 1$\sigma$ error of the emission peak (and all the other spot parameters) by fitting a Gaussian curve to the histogram plots. 
The centre of the peak height distribution (0.064 $\pm$ 0.015) for $\gamma$  = -30 km s$^{-1}$ is different from zero at the 4 - 4.5$\sigma$ level, indicating a significant detection (99.99\% confidence). 

It can be shown using the same method that the bright spot in the upper left velocity quadrant is also significant at the 4.4$\sigma$ level, and likely originates from the stream-disc impact region. A third faint spot occurs on the negative $V_y$-axis, with a marginal significance level of 3.2$\sigma$, the location suggests that it could be part of a broken-ring-like feature corresponding to the accretion disc. 
The significance levels of the rest of the spots are all below the 2$\sigma$ threshold.
In summary, although 
several spot-like features are present in the noisy \ion{N}{iii} $\lambda$4640/combined Bowen maps,
\textbf{only one sharp feature (statistically significant at the $>$ 4$\sigma$ level) lies on the positive $V_y$-axis and thus can be confidently identified as the donor signature in the case of J1814}. The availability of a robust absolute ephemeris in this case simplifies the search, but our methodology is appropriate even in cases such an ephemeris may not be available and thus all statistically significant components need to be considered.

\subsection{The Bowen blend diagnostic}   
\begin{figure}
	\includegraphics[width=\columnwidth]{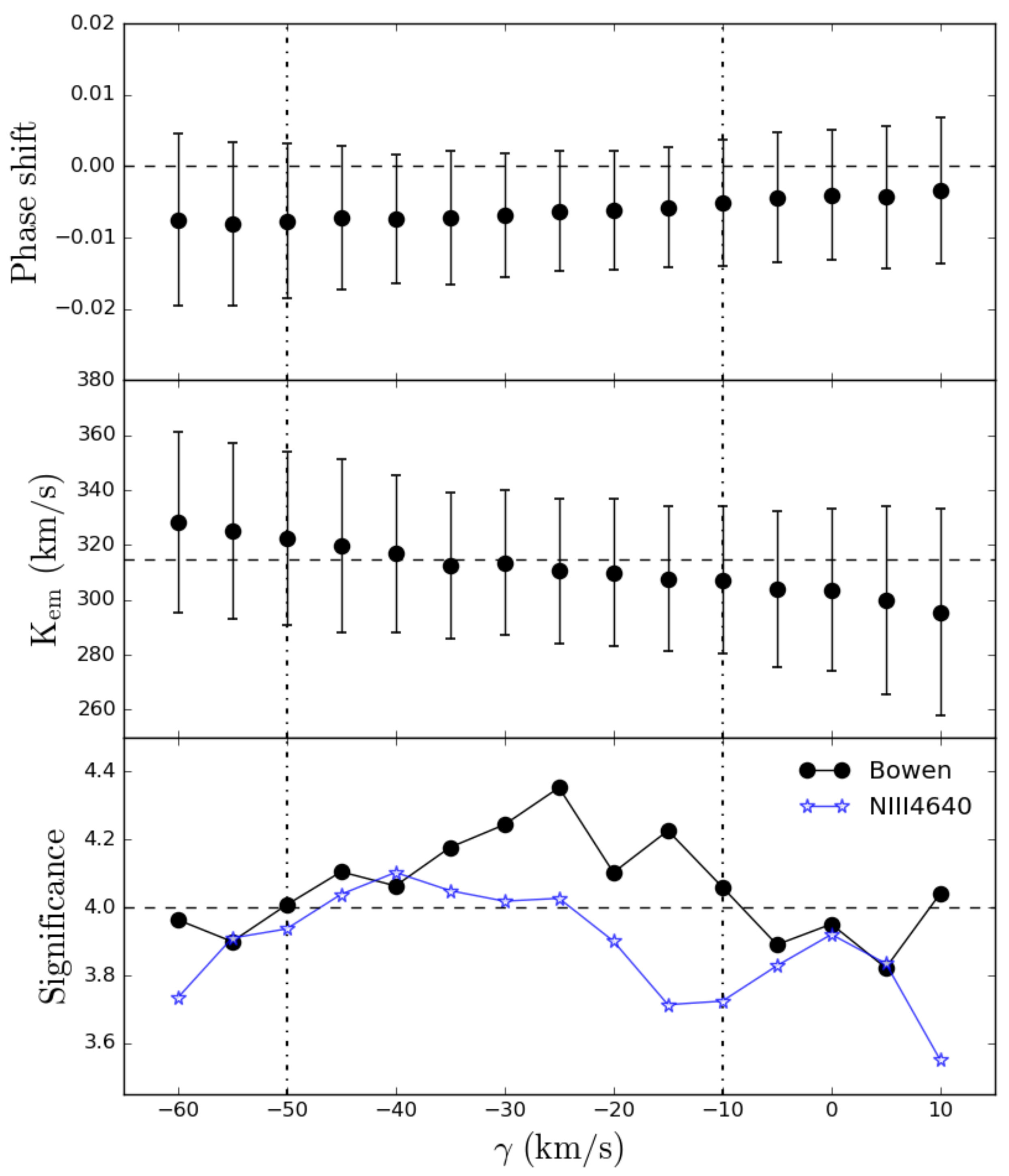}
    \caption{Diagram showing the best-fit solutions of the phase shift,$\rm~K_{em}$ amplitude with bootstrap estimate of 1$\sigma$ errors, and significance level of the donor spot feature derived from Bowen (black-filled circles) and \ion{N}{iii} $\lambda$4640 (blue stars) maps, as a function of the assumed systemic velocity. Vertical dash-dotted lines mark the preferred range of $\gamma$, over which the statistical significance of the secondary feature is greater than 4$\sigma$.}
    \label{fig:bowen_dd}
\end{figure}

Having confirmed that the Bowen emission is produced on the X-ray illuminated front side of the mass donor, and that the relevant emission feature is significant at a level higher than 4$\sigma$,
we aim to use the emission line diagnostics to obtain robust system parameter constraints. Along with the phase shift and peak emission, we show the distribution of the corresponding apparent radial velocity semi-amplitude$\rm~K_{em}$ ($<$ $K_2$) in Fig.~\ref{fig:histogram}. 
By fitting a Gaussian model to the histogram, we obtain$\rm~K_{em}$ = 313 $\pm$ 26 km s$^{-1}$. Similar analyses were applied to all the other \ion{N}{iii} $\lambda$4640 and the combined Bowen maps computed using the previously determined range of $\gamma$, and the results can be summarized in the form of a `\textit{Bowen diagnostic diagram}' (Fig.~\ref{fig:bowen_dd}).

\subsubsection{Systemic velocity}  
\label{sec:gamma}
In the new diagnostic diagram, we plot the phase shift,$\rm~K_{em}$ velocity (with 1$\sigma$ statistical errors) and the significance level of the combined Bowen spot feature as functions of the assumed $\gamma$. 
Due to the low signal-to-noise ratio of our data, we did not find a satisfactory $\gamma$ range that would minimise the reduced $\chi^{2}$ of a series of reconstructions (sensitivity to $\gamma$ is weak). Nevertheless, the map reveals a real donor signature, mapped from one or more sharp `S-wave' component(s) present in the trailed spectra, we can therefore expect that the best estimate of $\gamma$ would yield the maximum significance of the secondary feature.

From the bottom panel of Fig.~\ref{fig:bowen_dd}, it is clear that the significance of the combined Bowen spot is consistently higher than that of the \ion{N}{iii} $\lambda$4640 spot, signifying the merit of assigning multiple Bowen transitions to one image. In both cases, the level of significance peaks near the mid-point of the $\gamma$ range. As a final step in constraining the systemic velocity, we select a range (indicated by vertical dash-dotted lines) over which the significance of the Bowen spot is greater than the 4$\sigma$ level (horizontal dashed line). 
Hence, we use the combined Bowen spot and focus only on this final range of $\gamma$ ($\gamma=-30$ $\pm$ 20 km s$^{-1}$) for deriving the radial velocity semi-amplitude.

\subsubsection{$\mathrm{K_{em}}$ velocity}
The middle panel of Fig.~\ref{fig:bowen_dd} shows that within the preferred $\gamma$ range,$\rm~K_{em}$ amplitude is very weakly dependent on the underlying $\gamma$ assumption, with a maximum of 8 km s$^{-1}$ drift around the central value of 315 km s$^{-1}$ (marked by a horizontal dashed line). We therefore conclude that the systematics due to uncertainties in the main assumption underlying the tomography-based method can be easily tracked down and quantified. At low SNRs, it is clearly the case that the systematic uncertainty 
due to our assumption of $\gamma$ is subdominant.
Using the information from the diagram, we derive the best estimate of $\rm~K_{em}$ = 315 $\pm$ 28 (statistical) $\pm$ 8 (systematic) km s$^{-1}$; and $\Delta$$\mathrm{\phi_{spot}}$ = -0.006 $\pm$ 0.01 (statistical) $\pm$ 0.001 (systematic). 

The strategy described above can be easily extended to mid- to high-SNR cases, as well as cataclysmic variable (CV) systems, where a different emission line diagnostic is used \citep[e.g. \ion{Ca}{ii;}][]{2010MNRAS.401.1857V}. 
We will present a reanalysis of the higher SNR Bowen data of other previously published systems \citep{2008AIPC.1010..148C} with updated system parameter constraints in a future paper.
Assuming the emission components coming from the irradiated companion are detected, relevant constraints on true $K_2$ as well as the binary mass ratio $q$ can be inferred by applying the `K-correction' to the observed $\mathrm{K_{em}}$ amplitude. 

\section{Discussion}
\label{sec:discussion}

\subsection{The K-correction}
\citet{2005ApJ...635..502M} (hereafter, MCM05) 
showed that the deviation between the reprocessed light centre and the centre of mass of the Roche-lobe-filling donor$\rm~(K_{c}$ =$\rm~K_{em}/K_{2}$ $<$ 1) has a very weak dependence on the inclination angle $i$, but is strongly dependent on the mass ratio $q$, and the disc flaring angle $\alpha$ 
(see MCM05, figure 4).
With little or no knowledge of the disc shielding parameter, the K-correction is constrained between $\alpha$ = 0$\degr$ (maximum displacement) and the limit set by emission from the limb of the irradiated region (minimum displacement): $\rm~K_{em}/K_{2}$ $<$ 1 - 0.213 $q^{2/3}$$(1+q)^{1/3}$.

Using the new analysis toolset, we could also confirm the significance (4.2 - 4.7$\sigma$ level) of the \ion{He}{ii} $\lambda$5411 spot, and the phasing constraint ($\Delta$$\mathrm{\phi_{spot}}$ = 0.009 $\pm$ 0.009 $\pm$ 0.002) points to the inner Roche lobe of the donor star as a possible reprocessing site. 
However, we derived a velocity semi-amplitude significantly higher (3.3$\sigma$) than the Bowen$\rm~K_{em}$ velocity:
$\rm~K_{em, \ion{He}{ii}}$ = 477 $\pm$ 29 $\pm$ 27 km s$^{-1}$. 
We initially considered whether the \ion{He}{ii} $\lambda$5411 emission component could also come from the irradiated companion, albeit a different region of the Roche lobe - possibly due to differential shielding by the accretion disc.  
We then used the numerical solution for$\rm~K_c$ ($\alpha$ = 0$\degr$) derived in MCM05, for the lower inclination angle case (deduced from the lack of X-ray eclipses), to perform a K-correction for both$\rm~K_{em}$ velocities. 
\begin{figure}
	\includegraphics[width=\columnwidth]{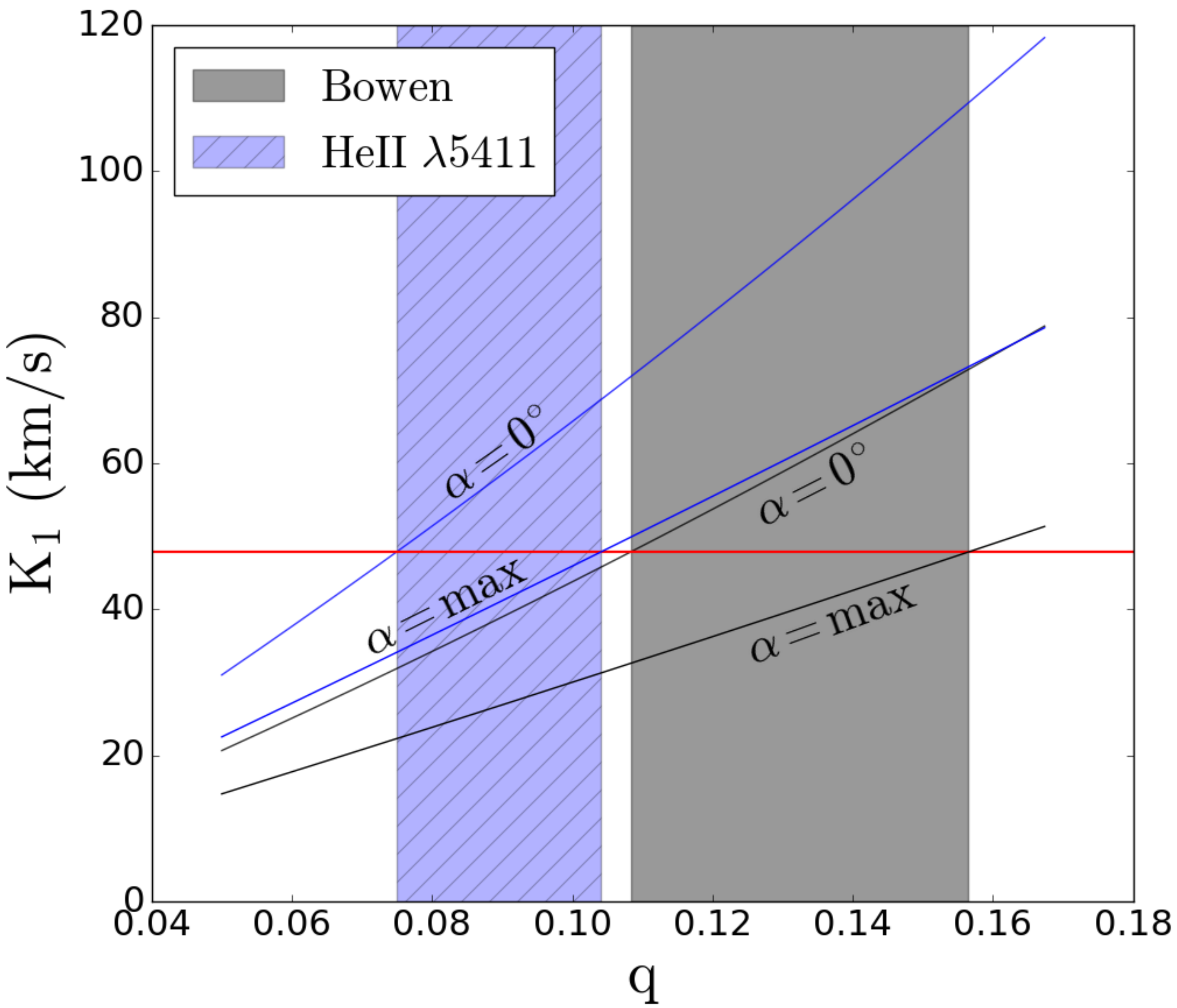}
    \caption{Constraints on $q$ using the K amplitude derived from the Bowen (grey shaded region) and \ion{He}{ii} $\lambda$5411 spot (blue shaded region). 
The horizontal red line denotes the estimate of $K_1$ from the pulsar solution.}
    \label{fig:K1_q_space}
\end{figure}

As the velocity amplitude of the neutron star ($K_1$ = 47.848 $\pm$ 0.001 km s$^{-1}$) is accurately known from the pulsar solution, the upper and lower limits for $q$ inferred from the Bowen and the 5411 $\AA$ spot can be easily visualized in a $K_1$ - q plane (Fig.~\ref{fig:K1_q_space}).
Not surprisingly, the two significantly different radial velocities with their associated errors (by adding the statistical and systematic uncertainties in quadrature) had led to very different constraints of $q$ (denoted by grey and blue shaded regions for the Bowen and \ion{He}{ii} $\lambda$5411 spot, respectively). 
Since there is no overlap between these shaded areas, there exists no value of $q$ that would satisfy both velocity constraints at the same time, i.e., only one of the two emission spots can originate from the surface of the donor. 
We must therefore focus on the $\rm~K_{em}$ velocity which most certainly traces the orbit of the companion to derive a feasible set of parameter constraints.

\begin{figure}
	\includegraphics[width=\columnwidth]{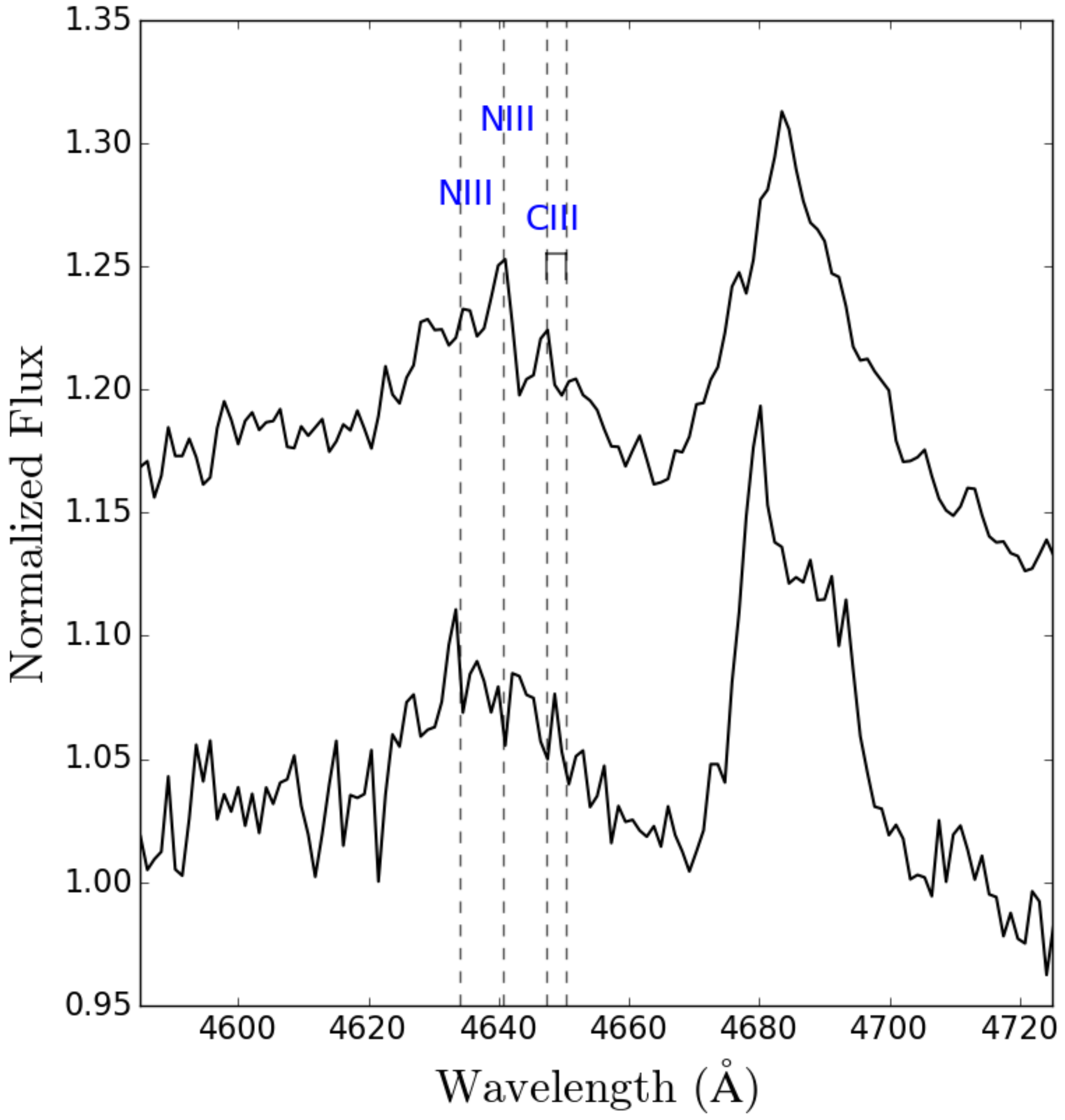}
    \caption{The Bowen/\ion{He}{ii} region of the average spectrum in the rest frame of the companion (\textit{top}) compared with a straight average spectrum (\textit{bottom}). At least three narrow peaks corresponding to known \ion{N}{iii}/\ion{C}{iii} transitions (vertical dashed lines) emerge in the Doppler corrected average. Top spectrum was calculated using the $\mathrm{K_{em}}$ amplitude and the systemic velocity values reported in Table~\ref{tab:allparams}.}
    \label{fig:spec_dop_corrected}
\end{figure}

Previous studies have shown that the irradiated atmospheres of the companion star in many LMXBs are powerful emitters of the fluorescence \ion{N}{iii} components. 
Since we have verified that narrow lines are also present in the Bowen region of J1814, we may, by extension, attribute them to the donor. 
A further support to this hypothesis is provided by the correct phasing and the $>$4$\sigma$ significance level of the relevant spot feature in the reconstructed Doppler map. 
Similar to most of the other Bowen targets, the strongest component around 4630-4650 $\AA$ is \ion{N}{iii} $\lambda$4640, which only becomes visible in the Doppler corrected average spectrum (Fig.~\ref{fig:spec_dop_corrected}). 
Therefore we are confident that the grey region (0.108 < $q$ < 0.157) in Fig.~\ref{fig:K1_q_space} represents a reliable range of the mass ratio for J1814.
Following this argument, we rule out the possibility that the component detected in the 5411 $\AA$ emission profile comes from the X-ray illuminated front face of the companion.
A tentative scenario for the origin of \ion{He}{ii} $\lambda$5411 emission is proposed in section \ref{sec:spider}. 

\subsection{Binary parameter estimation}
\label{sec:parameters}
The detection of Bowen emission lines in J1814 opens up a special opportunity to constrain the radial velocity of the companion star ($K_2$), and thereby the mass of the accreting millisecond pulsar ($M_1$) via the mass function equation:
\begin{equation}
f(M) = \frac{K_2^3\rm~P_{orb}}{2\pi G} = \frac{M_1\rm~sin^3(i)}{(1+q)^2}.
\label{eq:massfunc}
\end{equation}
The observed lower limit to $K_2$ $\rm~(K_{em}$ = 315 $\pm$ 29 km s$^{-1}$), combined with the precise determination of $K_1$, delivers an absolute upper limit to the mass ratio of $q$ $<$ 0.167, which implies, using Paczy\'nski's approximation:\(\rm~sin(\alpha_{max}) \simeq R_2/a \simeq 0.462\rm~q^{1/3}/(1+q)^{1/3}\) \citep{1971ARA&A...9..183P}, that the disc opening angle $\alpha$ cannot be higher than $\sim$ 14$\degr$.

In order to arrive at the most reliable estimation of true $K_2$, we developed an algorithm to extend the MCM05 numerical solutions for the K-correction to a set of fast fourth-order polynomial approximations:
\begin{equation}
\label{eq:kcor_poly}
K_c = \frac{\rm~K_{em}}{K_2} \simeq N_0 + N_1q + N_2q^2 + N_3q^3 + N_4q^4,
\end{equation}
so that $\alpha$ can take any value between 0 and 14$\degr$. 
By interpolating over the grid of K-correction functions (for the $i$ = 40$\degr$ case and $\alpha$ from 0 to 18$\degr$ in steps of 2$\degr$) provided in \citet{2005ApJ...635..502M}, 
we can generate reasonable model K-corrections as functions of $q$ for any arbitrary disc opening angle within the given bounds.
The best fitting polynomial can be obtained with the generated model and applied to the measured$\rm~K_{em}$ amplitude. 

With the measurements of the binary orbital period$\rm~(P_{orb}$) and the projected radial velocity semi-amplitude $K_2$, equation~(\ref{eq:massfunc}) gives the minimum allowable mass of the compact primary $M_1$ (as sin($i$) $\lesssim$ 1 and 1+$q$ $>$ 1).
Precise determinations of stellar masses require knowledge of the mass ratio $q$ as well as the inclination angle $i$.

\subsubsection{Monte-Carlo analysis}
To fully account for uncertainties in all of the observables/input parameters, we performed a Monte Carlo analysis, and considered a random distribution of inclination angles to establish the probability density functions (PDF's) of the unknown system parameters. 
By selecting synthetic$\rm~K_{em}$, $K_1$ and$\rm~P_{orb}$ values from a Gauss-normal distribution (with a mean and 1$\sigma$ error equal to the observed values; see Table~\ref{tab:allparams}), 
a set of simulated data can be obtained at each Monte Carlo trial and used as inputs to equation~(\ref{eq:massfunc}). 
Since a firm lower limit on $\alpha$ for J1814 is not known, we chose to let $\alpha$ follow a uniform distribution between 0 and 14$\degr$. A value as low as zero is perhaps unlikely given the presence of an active accretion disk, but we prefer not to bias our results by adopting some ad-hoc estimate of $\alpha$. 
For every combination of randomly generated$\rm~K_{em}$, $K_1$ and $\alpha$ values ($\alpha$ is a continuous variable), we determined an accurate K-correction for the $\alpha$ angle using the previously developed approximation tool. The updated $K_c$ was then applied to$\rm~K_{em}$, followed by evaluation of the solution for the binary mass ratio ($q$ =$M_2$/$M_1$=$K_1$/$K_2$) and the corresponding true $K_2$. 

\begin{figure}
	\includegraphics[width=\columnwidth]{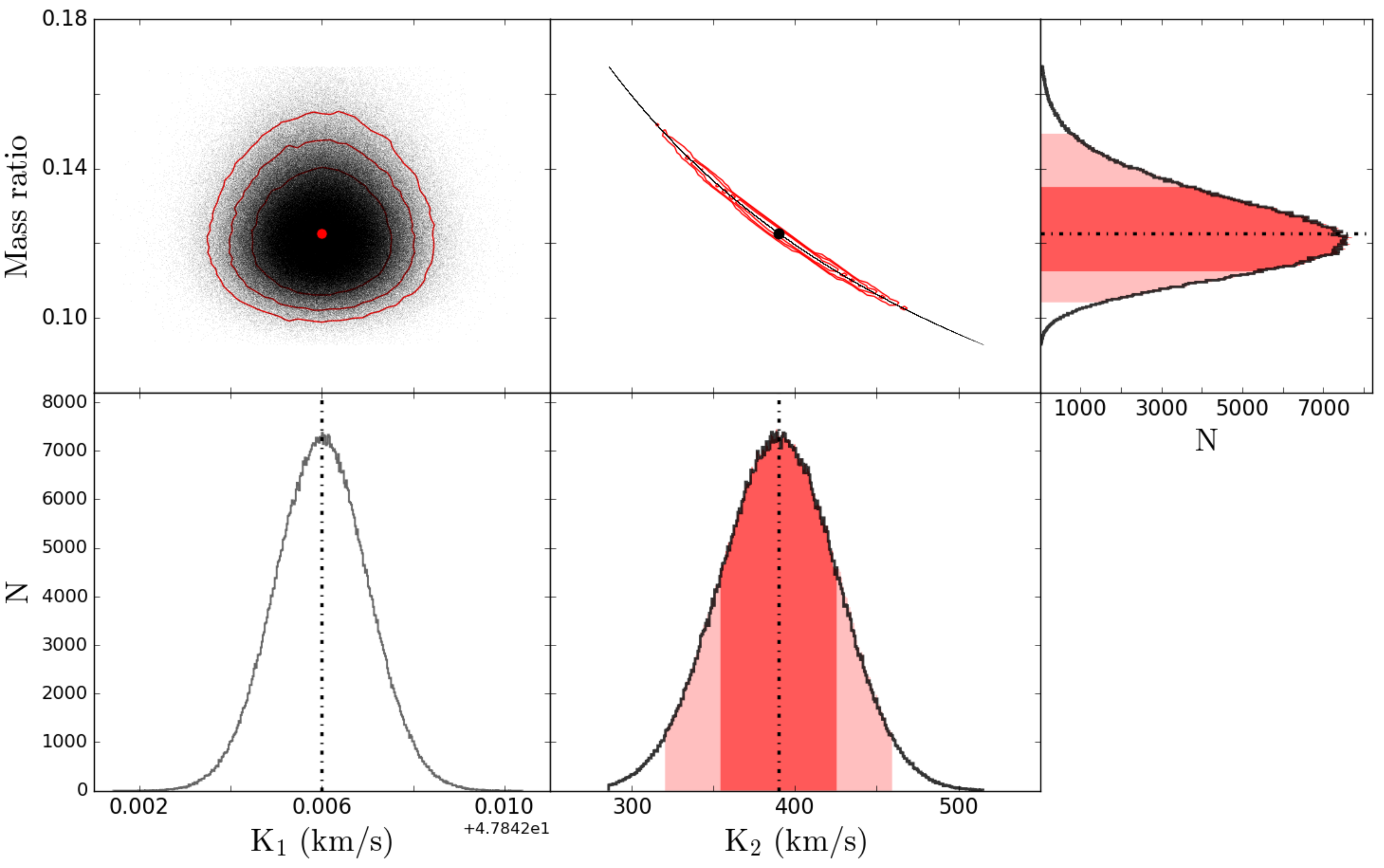}
   \caption{Probability density functions of $K_2$ and $q$ derived from our Monte Carlo analysis ($10^6$ trials) for 35$\degr$$\leq$ $i$ $\leq$ 78$\degr$. Dash-dotted lines denote the locations of the median ($\mathrm{50^{th}}$ percentile). Red-shaded areas correspond to the regions between the $\mathrm{2.5^{th}}$ and $\mathrm{97.5^{th}}$ percentiles. 68\% of the data fall within the darker regions. Best estimates of these system parameters and their associated 1$\sigma$ errors are given in Table~\ref{tab:allparams}.}
    \label{fig:q_K2_PDFs}
\end{figure}

\begin{figure}
	\includegraphics[width=\columnwidth]{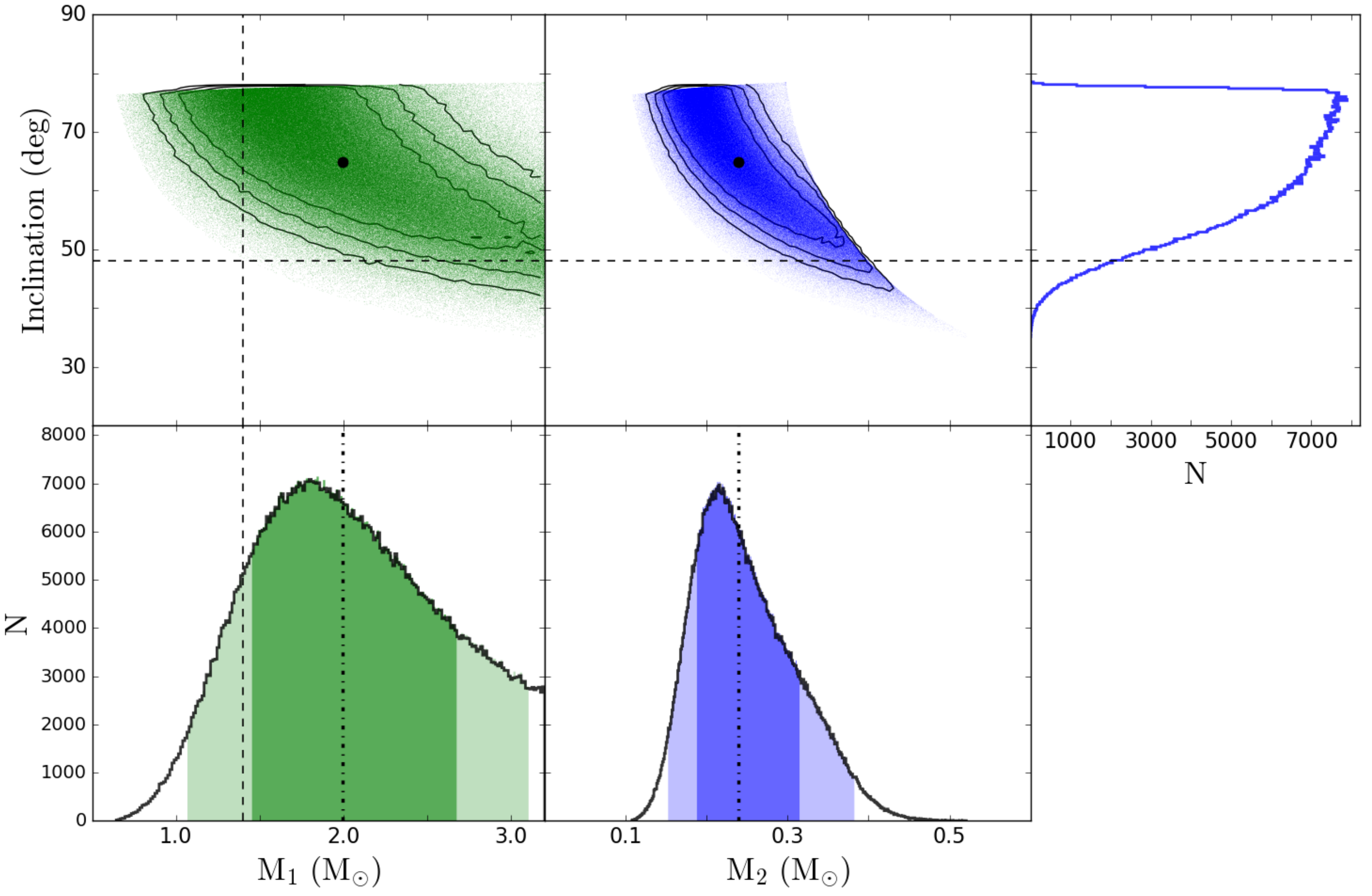}
   \caption{PDF's of the masses of the stellar components. The median values are denoted by dash-dotted lines. As in the previous PDF's, dark/light areas correspond to the regions between the $\mathrm{16^{th}}$/$\mathrm{2.5^{th}}$ and $\mathrm{84^{th}}$/$\mathrm{97.5^{th}}$ percentiles. The canonical value for the neutron star mass (1.4 $M_{\odot}$), denoted by a vertical dashed line, lies just outside the 68\% confidence limits. The horizontal dashed line denotes the lower limit to the inclination angle ($i$ $\gtrsim$ 48$\degr$) for an assumed pulsar mass of 1.4 $M_{\odot}$.} 
    \label{fig:M1_M2_MC}
\end{figure}

With a set of synthetic q, $K_2$ and $\mathrm{P_{orb}}$ in place, we proceeded to calculate a lower limit to the inclination angle by assuming a maximum stable neutron star mass of 3.2 $M_{\odot}$.
The upper limit to $i$ was simply estimated based on Paczy\'nski's relation given the absence of X-ray eclipses \citep{2005ApJ...627..910K}.
This would result in an overall very loose constraint on the inclination, 35$\degr$ $\lesssim$ $i$ $\lesssim$ 78$\degr$. 
Therefore, during the last step, we used a uniform distribution for cos($i$) (between the lower and upper bounds set by the particular combination of input values) to select a random inclination angle for deriving component masses. The same process was simulated $10^6$ times. Finally, the large number of separate outcomes evaluated in individual trials were assembled into probability density functions presented in Figures~\ref{fig:q_K2_PDFs} and~\ref{fig:M1_M2_MC}.

\subsubsection{The binary mass ratio and component masses}\label{qratio}
Perhaps the main disadvantage of the Bowen fluorescence technique lies in the fact that the measured$\rm~K_{em}$ velocity represents only a solid lower limit to $K_2$. 
However, in this case, we have the significant advantage of knowing the radial velocity semi-amplitude of the neutron star $K_1$ very accurately. 
As a result, both $K_2$ and $q$ PDF's (see Fig.~\ref{fig:q_K2_PDFs}) display near Gaussian distributions under the assumption of randomly distributed inclinations. 
By propagating the uncertainties in all estimated quantities - including especially the full uncertainty in the opening angle of the accretion disc and the systemic velocity - into the errors in the output quantities, we were able to derive robust estimates of the true velocity semi-amplitude of the donor star and the binary mass ratio
(with 1$\sigma$ errors of less than 10\% for both parameters): 
$K_2$ = $390^{+35}_{-36}$ km s$^{-1}$, 
$q$ = $0.123^{+0.012}_{-0.010}$. 

Additionally, a lower limit to the inclination for an assumed pulsar mass of 1.4 $M_{\odot}$ can be established ($i$ $\gtrsim$ 48$\degr$) from $K_2$ >$\rm~K_{em}$ = 315 $\pm$ 29 km s$^{-1}$.
The most secure constraint on $i$ (as shown in Fig.~\ref{fig:M1_M2_MC}) would be
35$\degr$ $<$ $i$ $<$ 78$\degr$ using the knowledge of lack of X-ray dips and combining Bowen$\rm~K_{em}$ velocity with $M_1$ < 3.2 $M_{\odot}$. 
This new inclination constraint thus provides a tighter lower bound compared with previously obtained limits based on burst oscillation properties \citep[26$\degr$ $<$ $i$ $<$ 50$\degr$;][]{2005ApJ...619..483B},  
or by assuming a magnitude limit (R $>$ 23.3) of the companion during quiescence \citep[22$\degr$ $<$ $i$ $<$ 77$\degr$;][]{2005ApJ...627..910K}. 

Fig.~\ref{fig:M1_M2_MC} shows the negative correlation between the inclination and the component masses, $M_1$ and $M_2$. 
The projected 1-dimensional PDF of the companion mass ($M_2$) has a tail extending to 0.38$\rm~M_{\odot}$ (95\% confidence) and a lower 95\% confidence limit of 0.15 $M_{\odot}$, which coincides with the minimum companion mass estimated from the pulsar mass function \citep{2003ATel..164....1M}.
The 68\% confidence interval (CI) for $M_2$ ($0.24^{+0.08}_{-0.05}$$\rm~M_{\odot}$) is also in excellent agreement with the $M_2$ constraints deduced from studies of the quiescent optical counterpart of J1814 \citep{2009A&A...508..297D,2013A&A...559A..42B}. 

The much wider $M_1$ PDF 
peaks beyond 1.4$\rm~M_{\odot}$ 
($2.0^{+0.7}_{-0.5}$$\rm~M_{\odot}$; 68\%). 
We note that the canonical mass of 1.4$\rm~M_{\odot}$ is not ruled out (1.1 $\lesssim$ $M_1$ $\lesssim$ 3.1; 95\%) with the lack of knowledge of $i$, 
and that a canonical NS mass would be more compatible with a relatively high inclination angle (63$\degr$ $\lesssim$ $i$ $\lesssim$ 76$\degr$; 68\%).
We already know that extremely high-mass neutron stars exist from the precise determinations of the mass of two radio millisecond pulsars, PSR J1614-2230 ($M_1$ = 1.928 $\pm$ 0.017 $\mathrm{M_{\odot}}$; \citealt{2010Natur.467.1081D}; see also \citealt{2016arXiv160300545F}) and PSR J0348+0432 ($M_1$ = 2.01 $\pm$ 0.04 $\mathrm{M_{\odot}}$; \citealt{2013Sci...340..448A}). 
An improvement in the measurement of the binary inclination is the key factor in achieving a tighter $M_1$ constraint for J1814, which may have the potential to exclude some soft equations of state.
A summary of system parameters with estimates of 1$\sigma$ uncertainties is given in Table~\ref{tab:allparams}. 
\begin{table}
	\centering
	\caption{XTE J1814-338 system parameters with estimates of 1$\sigma$/68\% uncertainties. $\mathrm{^a}$Adapted from \citep{2007MNRAS.375..971P}; $\mathrm{^b}$Based on \citep{2007MNRAS.375..971P}.}
	\label{tab:allparams}
	\begin{tabular}{ll} 
		\hline
                Parameter &  Value \\
		\hline
                $\mathrm{P_{orb}}$ $\mathrm{(d)^a}$ & 0.178110219(2) \\
                $T_0$$\rm~HJD(UTC)^a$ & 2452798.3539536(9) \\
                $\gamma$ (km s$^{-1}$) & -30 $\pm$ 20 \\
                $\mathrm{K_{em}}$ (km s$^{-1}$) & 315 $\pm$ 28 (statistical) $\pm$ 8 (systematic) \\
                $K_{1}$$(\rm~$km$\rm~s^{-1})^b$ & 47.848 $\pm$ 0.001 \\
                $K_{2}$ (km s$^{-1}$) & $390^{+35}_{-36}$ \\
                \noalign{\vskip 1mm}   
                $q$ ($M_2$/$M_1$) & $0.123^{+0.012}_{-0.010}$ \\
                $i$ ($\degr$) & 35 - 78\\ 
		\hline
                \noalign{\vskip 0.05mm}    
                $M_1$$\rm~(M_{\odot}$) & $2.0^{+0.7}_{-0.5}$ \\
                \noalign{\vskip 1mm}    
                $M_2$$\rm~(M_{\odot}$) & $0.24^{+0.08}_{-0.05}$\\
                \hline
	\end{tabular}
\end{table}

\subsection{The nature of the companion}
\label{sec:bloatedcompanion}
\begin{figure}
	\includegraphics[width=\columnwidth]{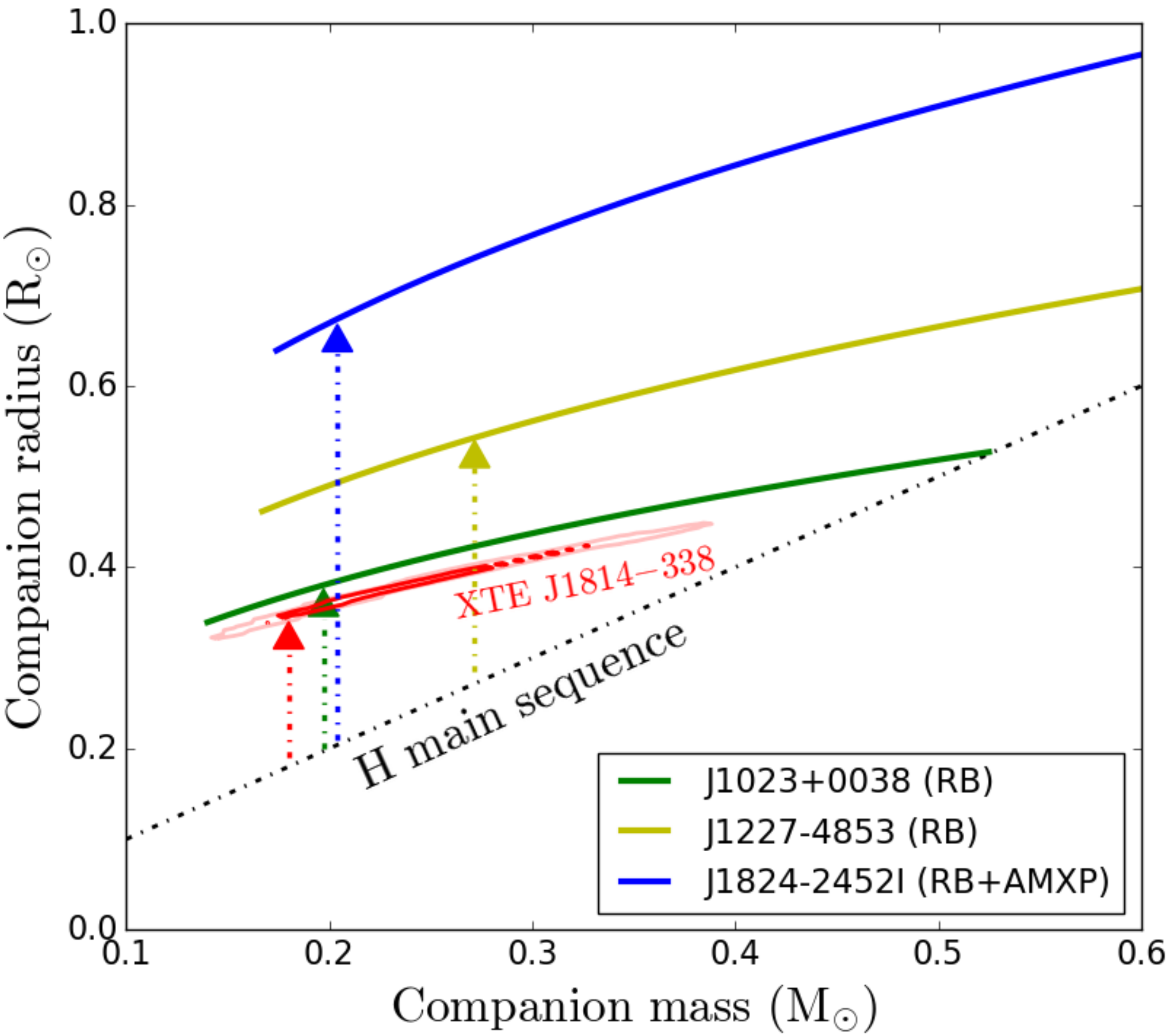}
   \caption{A comparison of the mass-radius relationships for the companion of J1814 and 3 t-MSPs \citep[so far all 3 are redbacks (RBs); see e.g.][]{2009Sci...324.1411A,2013Natur.501..517P,2014ApJ...790...39S,2014MNRAS.441.1825B} with the approximate relation for ordinary low-mass hydrogen main-sequence stars. The 68\% and 95\% confidence regions for J1814 (red contours) exclude the mass-radius relation for hydrogen main-sequence stars, indicating that the companion must be bloated. Dash-dotted lines with arrows denote nominal $M_2$ values determined for a pulsar mass of 1.4 $M_{\odot}$.}
    \label{fig:companion}
\end{figure}
Fig.~\ref{fig:companion} compares the mass-radius contours (showing the 68\% and 95\% confidence regions) for a Roche lobe-filling donor in a 4.275 hr binary (J1814) 
with the approximate relation for ordinary low-mass hydrogen main-sequence stars (using R/$\mathrm{R_{\odot}}$ = M/$\mathrm{M_{\odot}}$; black dash-dotted line). 
Our dynamical mass constraints confirmed previous suggestions of a main-sequence M-type companion in J1814, and that the companion is likely significantly bloated (given the 68\%/95\% confidence regions) presumably due to X-ray heating by an accretion-powered millisecond pulsar \citep[see also][]{2005ApJ...627..910K,2005ApJ...619..483B}. 

The fact that the radius of the secondary is 
$\approx$ 90\%
larger than expected (assuming $M_2$ = 0.18$\rm~M_{\odot}$ for a 1.4$\rm~M_{\odot}$ neutron star) is reminiscent of the first known transitional millisecond pulsar binary (t-MSP) PSR J1023+0038 \citep{2009Sci...324.1411A}. 
This archetypal redback system \citep[an eclipsing binary MSP with a companion of $\gtrsim$ 0.1 $M_{\odot}$;][]{2011AIPC.1357..127R} in the Galactic field has a similar orbital period to J1814 (green mass-radius curve in Fig.~\ref{fig:companion}) and a main-sequence-like, similar mass companion ($\sim$ 0.2$\rm~M_{\odot}$) that is bloated to 
$\approx$ 1.9
times the expected radius. 
Given these remarkable similarities, it is intriguing to speculate that before the 2003 outburst, the companion of XTE J1814-338 may have been irradiated by a rotation-powered (`recycled') \textbf{redback-like} pulsar which switches on during an X-ray quiescent state. Ablation processes driven by the pulsar spin-down power could bloat the secondary star significantly.

\subsection{A hidden `redback'?}
Evidence has been built over the past two decades to indicate strongly irradiated companions of several AMXP systems during quiescence. 
Optical light curves of SAX J1808.4-3658, IGR J00291+5934 and XTE J1814-338 all showed large-amplitude, sinusoidal modulations at the known orbital period \citep{2001MNRAS.325.1471H,2004ApJ...614L..49C,2007A&A...472..881D,2009A&A...508..297D}.
As first pointed out by \citet{2003A&A...404L..43B}, the irradiating luminosity necessary to account for the observed optical flux is about two orders of magnitude \textit{higher} than the quiescent X-ray luminosity $L_X$ ($\lesssim$ $10^{32}$ erg$\rm~s^{-1}$).  
On the other hand, the release of rotational energy via a relativistic particle wind (the spin-down luminosity; ;$\rm~L_{sd}$ $\sim$ $10^{34}$ erg$\rm~s^{-1}$) by a reactivated radio pulsar would be sufficient to illuminate the companion \citep{2008ApJ...675.1468H}.

From fitting multi-band, quiescent optical light curves of J1814 using a model for an irradiated secondary, \citet{2013A&A...559A..42B} noted an apparent discrepancy between the inferred value of $M_2$ (consistent with the companion being an M-type main-sequence star) and the day-side surface temperature of the companion ($\sim$ 5500 K), which is typical of an earlier spectral type (G or K-type) star \citep[see also][]{2009A&A...508..297D}. 
By assuming that the relativistic wind of an active radio pulsar irradiates the companion and increases its surface temperature, an estimate of the irradiation luminosity, $\mathrm{L_{irr}}$ = $\mathrm{\sigma_{SB}}$\big($\mathrm{T_{day}^4}$ - $\mathrm{T_{night}^4}$\big), can be related to the pulsar's spin-down energy as $\mathrm{L_{sd}}$ = 4$\pi$$a^2$$\mathrm{L_{irr}}$/$\mathrm{\epsilon_{irr}}$, where $\mathrm{\sigma_{SB}}$ is the Stefan-Boltzmann constant, $a$ is the orbital separation and $\mathrm{\epsilon_{irr}}$ is the irradiation efficiency. For an efficiency in the range of 0.1 to 0.3 typical of irradiated black widow ($M_2$ $<<$ 0.1 $M_{\odot}$) and redback MSP systems \citep{2013ApJ...769..108B}, the derived value of $\mathrm{L_{sd}}$ ($\sim$ [6 - 17] $\times$ $10^{34}$ erg $\mathrm{s^{-1}}$) is consistent with the spin-down luminosity of a 3.2 ms pulsar \citep{2013A&A...559A..42B}. 
This finding, combined with the new dynamical mass constraints and the bloated nature of the companion (see section \ref{sec:bloatedcompanion} and Fig.~\ref{fig:companion}), provides indirect evidence of a redback MSP during quiescence in XTE J1814-338 (although no evidence of pulsed radio emission of this source has been reported to date). 

The presence of a redback MSP might shed some light on the emission spot seen in the Doppler map of \ion{He}{ii} $\lambda$5411 in Fig.~\ref{fig:dopmaps}.
Since its $V_y$-coordinate is larger than the upper limit to $K_2$ derived from the \ion{N}{iii} $\lambda$4640/Bowen spot, it cannot be produced on the irradiated surface of the donor. 
Noticeable differences between the shape (or FWHM) of the \ion{He}{ii} $\lambda$5411 and the \ion{N}{iii} $\lambda$4640 spot suggest that the former might instead arise from a more extended structure, perhaps an intra-binary shock caused by the pulsar wind interacting with the ablated material from the donor.   
A puzzling behaviour that may be explained by the presence of a shock front was also noted by \citet{2009A&A...508..297D}. Quiescent optical light curves of J1814 exhibit a decrease in the V-band flux between phase 0.05 and 0.17.
The deep minimum around phase 0 (inferior conjunction of the donor star) is likely caused by a shock front eclipsing a residual accretion disc if present during quiescence, as suggested in \citet{2013A&A...559A..42B}.

\label{sec:spider}









\section{Conclusions}

We have presented the first detection of the donor star in the AMXP XTE J1814-338 using Doppler tomography of the Bowen region ($\lambda$$\lambda$4630-50) inspired by our earlier studies of Sco X-1 \citep{2002ApJ...568..273S} as well as a number of LMXBs in active states \citep{2008AIPC.1010..148C}.
The reconstructed Bowen Doppler map reveals a compact spot at phase 0 and $V_y$ $\sim$ 315 km s$^{-1}$, which has been interpreted as possible evidence of emission from the irradiated donor. 
In the lowest SNR regime, we used a bootstrap Monte-Carlo test to quantify the significance of the Bowen spot from ensembles of maps. 
The spot is statistically significant (> 4$\sigma$), and at the expected position of the donor star, thus providing a unique opportunity to constrain the radial velocity semi-amplitude of the companion of an AMXP. 

Since the Bowen components originate from the front face of the donor, the velocity semi-amplitude ($\mathrm{K_{em}}$) derived from the Bowen spot must be biased towards lower values. 
Based on the numerical solutions for $K_c$ = $\mathrm{K_{em}}$/$K_2$ \citep{2005ApJ...635..502M}, we developed an algorithm to compute synthetic K-corrections as functions of the mass ratio $q$ for any arbitrary disc opening angle $\alpha$ between 0 - 18$\degr$, which can be applied to $\mathrm{K_{em}}$ to get true $K_2$.
To account for uncertainties in all of the input parameters, we performed a Monte-Carlo analysis and considered a random distribution of inclination angles to establish the PDF's of the binary parameters. Under the most conservative scenario, we have the following constraints: $q$ = $0.123^{+0.012}_{-0.010}$; 35$\degr$ $<$ $i$ $<$ 78$\degr$; 1.5 $M_{\odot}$ $<$ $M_1$ $<$ 2.7 $M_{\odot}$; 0.19 $M_{\odot}$ $<$ $M_2$ $<$ 0.32 $M_{\odot}$ (68\%). 

The new dynamical mass constraints confirmed previous suggestions that the companion is a significantly bloated, main-sequence M-type star. Furthermore, a VLT campaign carried out between 2004 - 2009 (during quiescence) gave indications of a strongly irradiated companion that is heated by the relativistic particle wind of a rotation-powered MSP \citep{2009A&A...508..297D,2013A&A...559A..42B}. The presence of a redback-like MSP ($M_2$ $\sim$ 0.2) might also explain the puzzling emission feature seen in the \ion{He}{ii} $\lambda$5411 Doppler map, which likely originates from an intra-binary shock created from the interaction between the pulsar wind and the ablated material from the companion.

\section*{Acknowledgements} 
DS and TRM acknowledges support from the STFC under grant ST/L000733.
JC and TMD acknowledge support by the Spanish Ministerio de Economia y competitividad (MINECO) under grant AYA2013-42627.
JC acknowledges support by DGI of the Spanish Ministerio de Educaci\'on, Cultura y Deporte under grant PR2015-00397, also to the Leverhulme Trust through grant VP2-2015-04.
Based on observations collected at the European Organisation for Astronomical Research in the Southern Hemisphere under ESO programme ID 071.D-0372(A).




\bibliographystyle{mnras}
\bibliography{LW_reflist}








\bsp	
\label{lastpage}
\end{document}